\documentclass[usenatbib,referee]{mn2e}
\usepackage{graphicx}
\title{Optical multi-band observations of BL Lacertae during the outburst of 2005}
\author[S. M. Hu et al.]{Shao Ming Hu$^{1,2,3}$, J. H. Wu$^{1}$, G. Zhao$^{1,2}$\thanks{Corresponding author. E-mail: gzhao@bao.ac.cn} and X. Zhou$^{1}$ \\
       $^{1}$National Astronomical Observatories, Chinese Academy of Sciences, A20 Datun Road, Beijing, 100012, China\\
       $^{2}$Department of Space Science and Applied Physics, Shandong University at Weihai, 180 Cultural West Road, Weihai,
       \\ Shangdong, 264209, China\\
       $^{3}$Graduate School of Chinese Academy of Sciences, Beijing, 100049, China}

\begin{document}
\date{Received date  / Accepted date}

\pagerange{\pageref{firstpage}--\pageref{lastpage}} \pubyear{2006}

\maketitle

\label{firstpage}

\begin{abstract}
The aim of our observations is to investigate the intranight
variability properties and the spectral variability of BL
Lacertae. 799 optical multi-band observations were intensively
made with the Beijing-Arizona-Taiwan-Connecticut (BATC) 60/90cm
Schmidt telescope during the outburst composed of two subsequent
flares in 2005. The second flare, whose rising phase lasted at
least 44 days, was observed with amplitudes of more than 1.1 mag
in three BATC optical bands. In general, on intranight timescale
the amplitude of variability and the variation rate are larger at
the shorter wavelength, and the variation rate is comparable in
the rising and decaying phases within each band. A possible time
lag between the light curves in bands \emph{e} and \emph{m},
around 11.6 minutes, was obtained. Based on the analysis of the
colour index variation with the source brightness, the variability
of BL Lacertae can be interpreted as having two components: a
``strongly-chromatic'' intranight component and a
``mildly-chromatic'' internight component, which may be the
results of both intrinsic physical mechanism and geometric
effects.
\end{abstract}

\begin{keywords}
galaxies: BL Lacertae objects: general -- BL Lacertae objects:
individual: BL Lacertae -- galaxies: photometry
\end{keywords}

\section{Introduction}

BL Lacertae (BL Lac) is the prototype of BL Lac objects(BL Lacs).
In the unified scheme of radio-loud active galactic nuclei (AGNs),
BL Lacs and flat-spectrum radio quasars (FSRQs) are usually
collectively termed as blazars. BL Lacs are characterized by
non-thermal continuum emission across the whole electromagnetic
spectrum with absent or weak emission and absorption lines
\citep[e.g.][]{stickel}, variable and high polarization
\citep[e.g.][]{angel,impey,gabuzda}, large amplitude and rapid
variability at all wavelengths from radio to $\gamma$-rays (e.g.
\citealp{ravasio}; \citealp{bottcher2003}) and superluminal motion
of radio components (e.g. Denn, Mutel \& Marscher 2000). BL Lac is
a well-studied source that has been monitored in optical band for
more than a century. Variability on both long and short timescales
has been observed by many authors (Villata et al.
\citealp{villata2004a}, and references therein). Microvariability
has been detected since the early work by Racine \cite{racine}.
Intranight optical variability (INOV) of this object was studied
in detail during the long 1997 outburst. Nesci et al. \cite{nesci}
reported multi-band INOV based on the observations made in July
1997. The source was never found to be stable during their
campaign. They found that the amplitude of flux variation was
always larger at shorter wavelengths and no apparent time lags
among the light curves in different wavebands could be detected.
Speziali \& Natali \cite{speziali} monitored BL Lac in three
optical bands on three nights in August 1997. Their results
involved INOV on timescale ranging from 30 minutes up to 3.5
hours. They concluded that the magnitude of variability was larger
at shorter wavelengths and the source showed a marked trend to
become bluer when brighter. This chromatism of INOV was confirmed
by many other authors \citep[e.g.][]{matsumoto,ghosh, clements}.
Papadakis et al \cite{papadakis} analyzed the INOV, the timescales
of rising and decaying processes, and the time lags between
different optical bands. Their results showed that the variability
amplitudes increased with increasing frequency. The rising and
decaying timescales were comparable within each band, but
increased with decreasing frequency. The time lag between the
light curves in bands \emph{B} and \emph{I} was less than
$\sim\pm$0.4 hours. They also detected significant spectral
variations, which had the trend to become harder (bluer) as the
flux increased.

In order to investigate the details of the physical processes and
geometric conditions based on the brightness and spectral
variability, BL Lac has been intensively monitored in four
campaigns (see \citealp{ravasio}; \citealp{villata2002};
\citealp{bottcher2003}; Villata et al. \citealp{villata2004a};
Villata et al. \citealp{villata2004b}) by the Whole Earth Blazar
Telescope (WEBT) collaboration since 1999 to 2003. Most of these
optical multi-band data were collected and analyzed by Villata et
al. \cite{villata2004a}. Their results from the analysis of colour
index revealed that the variability had two components, long term
(a few-day timescale) ``mildly-chromatic'' variations and strong
bluer-when-brighter chromatic short term variations (intraday
flares). Stalin et al. \cite{stalin} showed that BL Lac became
bluer when brighter on intranight timescale, while this trend was
less significant on internight timescale. Their results may also
suggest that the variability of BL Lac has two components. They
found that the dependence of the spectral variability on its
brightness was different from that of S5 0716+714. On the
contrary, some objects were reported to become redder when
brighter. For example, Ramirez et al \cite{ramirez} presented that
the spectrum of PKS 0736+017 became redder when brighter on both
internight and intranight timescales. This different property may
be introduced by different physical processes or physical
environments.

To shed more light on the INOV and spectral variation with respect
to flux variation on both internight and intranight timescales, we
carried out dense observations on BL Lac. We describe the
observations and data reduction in the following section. Sect.~3
gives the results of our observations. A summary is presented in
Sect.~4.

\section{Observations and data reduction}

Our observations were carried out with a 60/90cm f/3 Schmidt
telescope, which is located at the Xinglong Station of the
National Astronomical Observatories of China. The telescope is
equipped with a 15-colour intermediate-band photometric system,
covering a wavelength range of 3\,000--10\,000 \AA. A Ford
Aerospace $2\,048\times2\,048$ CCD camera is mounted at its main
focus. The CCD has a pixel size of 15 microns and its field of
view is $58'\times58'$, resulting in a resolution of
1.7\arcsec/pixel. The telescope and the photometric system are
mainly used to carry out the BATC survey \citep{zhou2005}.

We performed optical observations in the BATC \emph{e}, \emph{i}
and \emph{m} bands in a cyclic mode. Their central wavelengths are
4885, 6685 and 8013 \AA, respectively. For good compromise between
photometric precision and temporal resolution, the exposure times
were 150 seconds in the BATC \emph{i} band and mostly 240 seconds
in the \emph{e} and \emph{m} bands. Because a frame size of
$512\times512$ pixels ($14\farcm5\times14\farcm5$) is large enough
to cover BL Lac and its comparison stars, we only read out the
central $512\times512$ pixels of the CCD images. The readout time
is about 5.6 seconds. So we can achieve a temporal resolution of
about 12 minutes in each band. This quasi-simultaneous
measurements enable us to investigate the colour variation. BL Lac
was observed on 26 nights during the period from July 5, 2005 to
November 18, 2005.

All images were processed with the automatic data reduction
procedure including bias subtraction, flat fielding, position
calibration, aperture photometry and flux calibration. The adopted
aperture radius was 4 pixels (6.8\arcsec), since the average FWHM
of the object and the comparison stars was 3.5\arcsec. The inner
and outer radii of the sky annulus were set as 6 and 9 pixels
(10.2\arcsec and 15.3\arcsec), respectively. The comparison stars
B, C and H from Smith et al. \cite{smith} were used for flux
calibration. Their BATC magnitudes were obtained by observing them
and three BATC standard stars, HD~19445, HD~84937 and BD+17d4708,
on a photometric night. They were listed in Table~\ref{batcstd}.
The star ID is followed by its BATC \emph{e}, \emph{i}, \emph{m}
magnitudes and errors in Table~\ref{batcstd}. The BATC magnitudes
of BL Lac can be obtained by differential photometry. The BATC
magnitudes can also be transferred to standard Johnson-Cousins
magnitudes by the relations obtained by Zhou et al.
\cite{zhou2003}. The details on data reduction procedure were
described by Yan et al. \cite{yan} and Zhou et al.
\cite{zhou2003}. Our photometry results are listed in
Tables~\ref{e}--\ref{m}. These three tables have the same format.
The universal date of observations (Col.~1) is followed by the
universal time at the middle of exposure (Col.~2), Julian date at
the middle of exposure (Col.~3), exposure time (Col.~4), BATC
magnitudes (Col.~5), photometry error (Col.~6) and the difference,
$\delta_{x}$ (it is the difference between the differential
magnitude of comparision stars B, C $(B_{x}-C_{x})$, and the
average of $(B_{x}-C_{x})$. \emph{x} indicates the BATC band
\emph{e}, \emph{i} or \emph{m}.), which indicates the confidence
of the observations (Col.~7). These three full tables are only
available in electronic form.

\begin{table}
\caption{BATC Magnitudes of the three comparison stars}             
\label{batcstd}      
\centering                          
\begin{tabular}{c c c c}        
\hline\hline                 
star & \emph{e}& \emph{i} &  \emph{m} \\
 ID  &  (mag) &  (mag) &  (mag) \\     
\hline                        
   B & 13.893$\pm$0.011 & 12.148$\pm$0.010 & 11.592$\pm$0.005 \\      
   C & 14.859$\pm$0.018 & 13.970$\pm$0.012 & 13.755$\pm$0.012 \\
   H & 15.229$\pm$0.025 & 13.913$\pm$0.013 & 13.509$\pm$0.010 \\
\hline                                   
\end{tabular}
\end{table}

\begin{table}
\caption{BATC \emph{e} magnitudes of BL Lac}             
\label{e}      
\centering                          
\begin{tabular}{c c c c c c r}        
\hline\hline                 
  UT Date  &  UT   &  JD   &  Exposure Time &\emph{e}&  $e_{err}$ & $\delta_{e}$ \\
yyyy/mm/dd & hh:mm:ss & (day) & (seconds) & (mag) & (mag)  & (mag)\\    
\hline                        
 2005/07/05 & 19:05:25 & 2453557.29541 &  240 & 16.126 & 0.032 & -0.003 \\
 2005/07/05 & 19:19:26 & 2453557.30518 &  240 & 16.079 & 0.032 &  0.003 \\
 2005/07/06 & 19:20:50 & 2453558.30615 &  300 & 16.239 & 0.050 &  0.000 \\
 2005/07/15 & 17:59:35 & 2453567.24976 &  240 & 16.011 & 0.078 &  0.029 \\
 2005/07/15 & 18:16:18 & 2453567.26123 &  240 & 16.008 & 0.064 &  0.036 \\
\hline                                   
\end{tabular}
\end{table}

\begin{table}
\caption{BATC \emph{i} magnitudes of BL Lac}             
\label{i}      
\centering                          
\begin{tabular}{c c c c c c r}        
\hline\hline                 
  UT Date  &  UT   &  JD   &  Exposure Time &\emph{i}&  $i_{err}$ & $\delta_{i}$ \\
yyyy/mm/dd & hh:mm:ss & (day) & (seconds) & (mag) & (mag)  & (mag)\\    
\hline                        
 2005/07/05 & 19:09:31 & 2453557.29834 &  150 & 14.840 & 0.012 & -0.011 \\
 2005/07/05 & 19:24:42 & 2453557.30884 &  150 & 14.866 & 0.012 &  0.011 \\
 2005/07/06 & 19:28:01 & 2453558.31104 &  180 & 14.982 & 0.016 &  0.000 \\
 2005/07/15 & 18:05:46 & 2453567.25391 &  150 & 14.762 & 0.018 &  0.014 \\
 2005/07/15 & 18:20:33 & 2453567.26416 &  150 & 14.818 & 0.018 &  0.024 \\
\hline                                   
\end{tabular}
\end{table}

\begin{table}
\caption{BATC \emph{m} magnitudes of BL Lac}             
\label{m}      
\centering                          
\begin{tabular}{c c c c c c r}        
\hline\hline                 
  UT Date  &  UT     &  JD   &  Exposure Time &\emph{m}&  $m_{err}$ & $\delta_{m}$ \\
 yyyy/mm/dd&hh:mm:ss & (day) & (seconds) & (mag) & (mag)  & (mag)\\
\hline                        
 2005/07/05 & 19:13:40 & 2453557.30127 &  240 & 14.258 & 0.013 &  0.003 \\
 2005/07/05 & 19:28:46 & 2453557.31152 &  240 & 14.245 & 0.014 & -0.003 \\
 2005/07/06 & 19:33:59 & 2453558.31519 &  300 & 14.410 & 0.018 &  0.000 \\
 2005/07/15 & 18:10:36 & 2453567.25732 &  240 & 14.205 & 0.017 & -0.005 \\
 2005/07/15 & 18:24:56 & 2453567.26733 &  240 & 14.205 & 0.017 & -0.010 \\
\hline                                   
\end{tabular}
\end{table}

\section{Results}

\subsection{Light curves}

\begin{figure}
   \centering
   \includegraphics[width=\columnwidth]{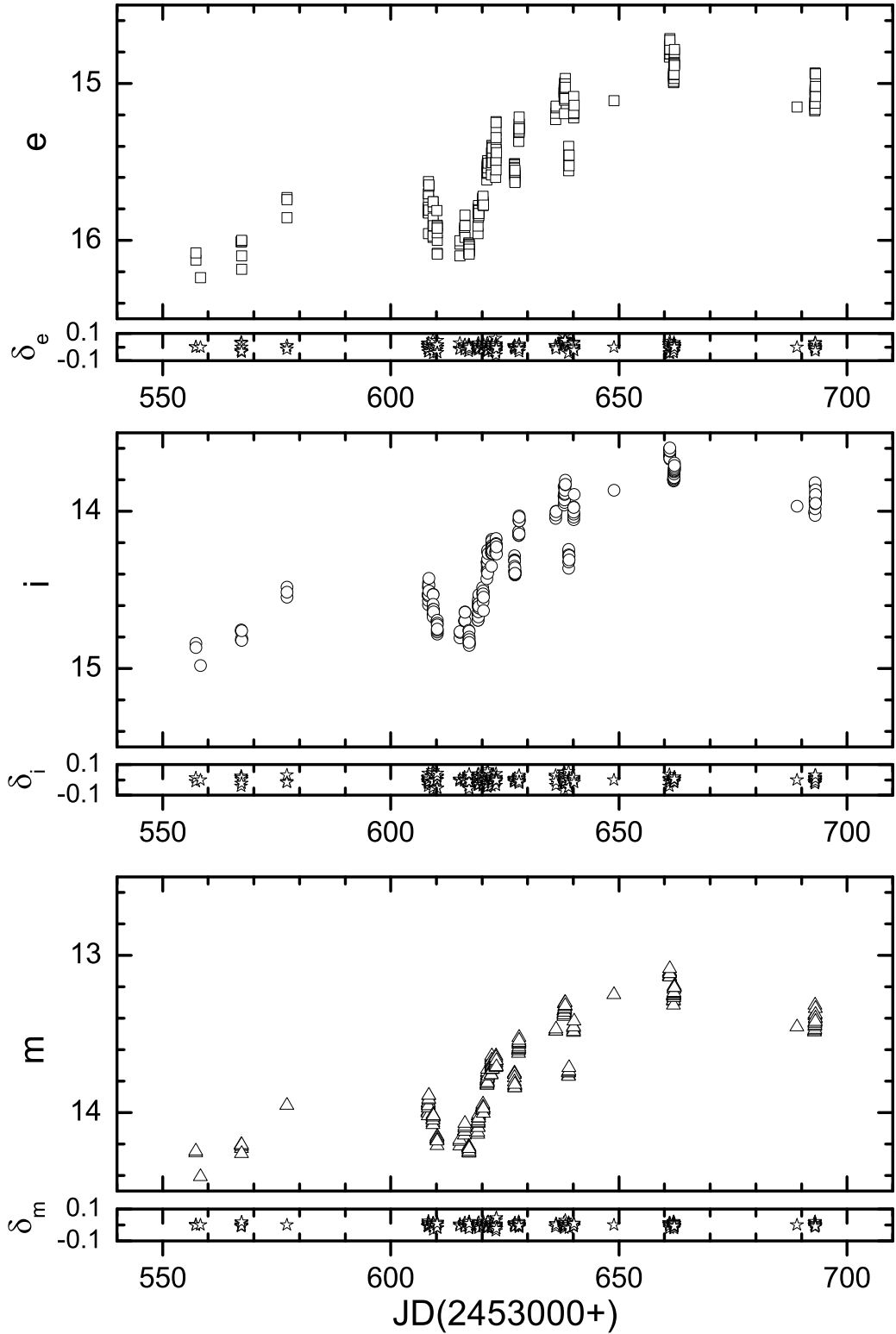}
      \caption{Light curves in bands \emph{e}, \emph{i} and \emph{m} from July 5, 2005 to November 18,
      2005. Squares, circles and triangles in the large panels
      represent the BATC \emph{e}, \emph{i} and \emph{m} magnitudes, respectively.
      Photometry
      errors are not plotted for clarity.
      Stars in the small panels represent the confidence of
      observations.
              }
         \label{lcall}
   \end{figure}

The light curves in bands \emph{e}, \emph{i} and \emph{m} are
displayed in Fig.~\ref{lcall}. Our photometry clearly shows that
BL Lac is violent and varies from day to day during the period of
our more than four months (JD from 2453557 to 2453693) monitoring.
One can see two flares apparently from Fig.~\ref{lcall}. The
rising phase of the second flare lasted 44 days (JD
2453617--2453661). The amplitudes of variation during the second
flare are 1.374, 1.252 and 1.174 mag in bands \emph{e}, \emph{i}
and \emph{m}, respectively. The variability amplitudes increased
with increasing frequency. It is a pity that we did not completely
record these two flares due to the weather conditions or other
campaigns.

One can see that many fast oscillations superimposed on the long
term variations. Our multi-band observations on many nights gave
us the opportunity to intensively investigate the intranight
variability. The criterion presented by Jang \& Miller \cite{jang}
was adopted to decide whether the source was variable or not. It
is defined as $C=\sigma_{T}/\sigma$, where $\sigma_{T}$ is the
standard deviation of the source magnitude, $\sigma$ are the
standard deviation of $\delta_{x}$. Because $\delta_{x}$ presents
the variation of the comparison stars, $\sigma$ indicates the
observational error that provides a more accurate determination of
the actual error than the normal photometric error. $C$ is taken
as the confidence level of variability. The source will be claimed
to be variable at 99\% confidence level when $C\geq2.576$. The
object is decided to be variable only when $C\geq2.576$ at least
in two bands if it is monitored in three or more bands (Jang \&
Miller \citealp{jang}; Stalin et al. \citealp{stalin}). To
quantitatively analyze the INOV, the intranight variability
amplitude is defined as \citep{heidt}:

\begin{equation}
      Y=\frac{100}{\langle D \rangle}\sqrt{(D_{max}-D_{min})^{2}-2\sigma^{2}}\%
\end{equation}
where $\langle D \rangle$, $D_{max}$ and $D_{min}$ are the
average, maximum and minimum of the magnitudes on one light curve.
$\sigma$ is the same as what is described above.

\begin{figure}
   \centering
   \includegraphics[width=\columnwidth]{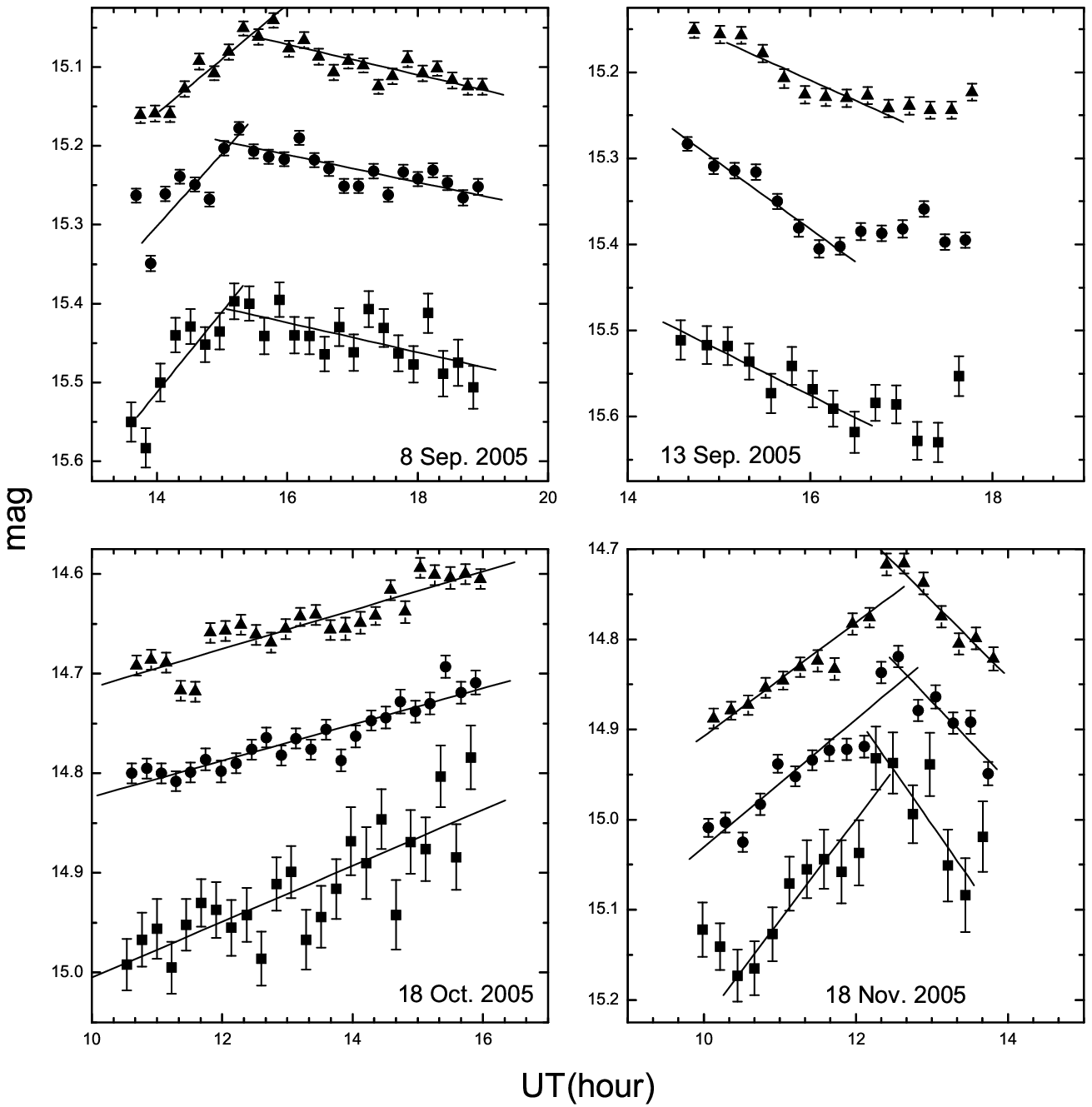}
      \caption{Examples of intranight light curves on four nights. Filled
      squares present the light curves in the BATC \emph{e} band. Filled
      circles and filled triangles present the light curves of
      $i+1.0$ and $m+1.4$, respectively (shift just for clarity). The solid lines show the
      linear fittings to the rising and decaying parts of lights
      curves.
              }
         \label{lcintra}
   \end{figure}

There are 19 out of 26 nights on which the data points are more
than 5 in each band. The confidence level of variability was
calculated for these 19 nights in each band. According to the
criterion described above, the object was variable on 9 nights.
The intranight variability amplitude, $Y$, was calculated for
these 9 ``variable nights''. The variability amplitude in band
\emph{e} is generally larger than those in bands \emph{i} and
\emph{m}. The average values are 1.22\%, 1.00\% and 0.82\%,
respectively. It is larger at the higher frequency. This is
consistent with the results of Nesci et al. \cite{nesci}, Speziali
\& Natali \cite{speziali} and Papadakis et al. \cite{papadakis}.
In order to compare the variation rates of the light curves in
different bands at both the rising and decaying phases, we fitted
the light curves on the 9 ``variable nights'' with a linear model.
The slope of the linear least square fitting is taken as the
variation rate. Fig.~\ref{lcintra} gives examples of linear
fittings on four nights. One can see from Fig.~\ref{lcintra} that
the light curves in three bands are consistent with each other.
All the analysis results are listed in Table~\ref{tintraday}, in
which the universal date of observations (Col.~1) is followed by
the number of frames in each band (Col.~2), the confidence level
of variability with the label of being variable or not (Col.~3,
V/N indicates the source is variable or not.), the intranight
variability amplitude (Col.~4), the variation rates and their
errors in the rising and decaying parts (Col.~5 and Col.~6). The
statistic results show that the variation rates of the light
curves in band \emph{e} is larger than those in bands \emph{i} and
\emph{m} both in the rising and decaying parts. It is larger in
the shorter wavebands in general. The average values of the
variation rate are 0.065, 0.057 and 0.042 in the \emph{e},
\emph{i} and \emph{m} bands for the rising parts, respectively,
while for the decaying phases, they are respectively $-$0.064,
$-$0.061 and $-$0.051. The variation rates are comparable in the
rising and decaying phases.

\begin{table}
\caption{Results of the intranight variability analysis, all dates are in the same year 2005.}             
\label{tintraday}      
\centering                          
\begin{tabular}{l l c c c c}        
\hline\hline                 
Date     & Band &C(V/N)  &Y  &$S_{A} \pm E_{A}$   &$S_{D} \pm E_{D}$       \\
mm/dd   &  (N)  &        &\% & $\times 10^{-2}$   & $\times 10^{-2}$       \\
\hline                        
07/15  &e(7)  &2.217(N)  &   &   &        \\
       &i(7)  &1.264(N)  &   &   &         \\
       &m(7)  &1.845(N)  &   &   &          \\
08/25  &e(14) &4.385(V)  &2.09   &7.7$\pm$4.3  &          \\
       &i(14) &1.909(N)  &1.15   &6.3$\pm$0.7  &          \\
       &m(14) &4.489(V)  &0.93   &3.4$\pm$0.6  &          \\
08/26  &e(9)  &4.262(V)  &   &   &          \\
       &i(9)  &1.221(N)  &   &   &          \\
       &m(9)  &1.446(N)  &   &   &          \\
08/27  &e(8)  &3.538(V)  &   &   &          \\
       &i(8)  &0.918(N)  &   &   &          \\
       &m(7)  &1.276(N)  &   &   &         \\
09/02  &e(6)  &2.853(V)  &0.88   &7.2$\pm$3.9  &          \\
       &i(6)  &3.187(V)  &0.39   &5.2$\pm$1.5  &          \\
       &m(6)  &2.655(V)  &0.51   &5.8$\pm$1.2  &          \\
09/03  &e(12) &1.785(N)  &   &   &          \\
       &i(12) &0.934(N)  &   &   &          \\
       &m(12) &1.136(N)  &   &   &          \\
09/05  &e(16) &3.349(V)  &1.12   &3.5$\pm$1.0  &         \\
       &i(16) &1.779(N)  &1.07   &4.0$\pm$0.3  &          \\
       &m(16) &4.008(V)  &0.91   &3.0$\pm$0.3  &          \\
09/06  &e(11) &1.325(N)  &   &   &          \\
       &i(11) &1.556(N)  &   &   &          \\
       &m(11) &1.831(N)  &   &   &          \\
09/07  &e(20) &1.687(N)  &   &   &          \\
       &i(20) &1.440(N)  &   &   &          \\
       &m(20) &2.859(V)  &   &   &          \\
09/08  &e(24) &3.343(V)  &1.21   &10.2$\pm$2.0  &$-1.9\pm$0.6   \\
       &i(24) &2.285(N)  &1.19   &9.2$\pm$0.7  &$-1.7\pm$0.2   \\
       &m(24) &4.105(V)  &0.87   &6.9$\pm$0.7  &$-1.9\pm$0.3   \\
09/09  &e(9)  &3.929(V)  &   &   &          \\
       &i(13) &1.760(N)  &   &   &          \\
       &m(13) &1.016(N)  &   &   &          \\
09/13  &e(14) &5.487(V)  &0.76   &   &$-5.3\pm$1.2   \\
       &i(14) &5.605(V)  &0.85   &   &$-7.7\pm$0.6   \\
       &m(14) &5.069(V)  &0.67   &   &$-4.8\pm$0.7   \\
09/14  &e(13) &2.712(V)  &1.00   &5.9$\pm$1.2  &        \\
       &i(13) &2.874(V)  &0.87   &5.8$\pm$0.4  &          \\
       &m(13) &3.047(V)  &0.73   &3.6$\pm$0.5  &          \\
09/24  &e(17) &2.349(N)  &   &   &          \\
       &i(17) &1.450(N)  &   &   &          \\
       &m(17) &2.063(N)  &   &   &          \\
09/25  &e(8)  &2.338(N)  &   &   &          \\
       &i(8)  &1.029(N)  &   &   &          \\
       &m(8)  &1.474(N)  &   &   &          \\
09/26  &e(8)  &1.964(N)  &0.88   &3.5$\pm$2.7  &          \\
       &i(8)  &5.477(V)  &1.15   &6.2$\pm$0.6  &          \\
       &m(8)  &2.800(V)  &0.53   &2.9$\pm$0.7  &          \\
10/17  &e(10) &1.384(N)  &   &   &          \\
       &i(10) &1.332(N)  &   &   &          \\
       &m(10) &1.624(N)  &   &   &          \\
10/18  &e(24) &3.306(V)  &1.41   &2.8$\pm$0.4  &         \\
       &i(24) &3.932(V)  &0.83   &1.8$\pm$0.1  &          \\
       &m(24) &2.992(V)  &0.93   &1.9$\pm$0.1  &          \\
11/18  &e(17) &4.507(V)  &1.59   &11.2$\pm$1.8 &$-12.1\pm$3.8   \\
       &i(17) &3.481(V)  &1.47   &7.1$\pm$0.4  &$-8.8\pm$1.3   \\
       &m(17) &5.289(V)  &1.28   &6.3$\pm$0.5  &$-8.5\pm$1.0   \\
\hline                                   
\end{tabular}
\end{table}

\subsection{Time lag}

\begin{figure}
   \centering
   \includegraphics[width=\columnwidth]{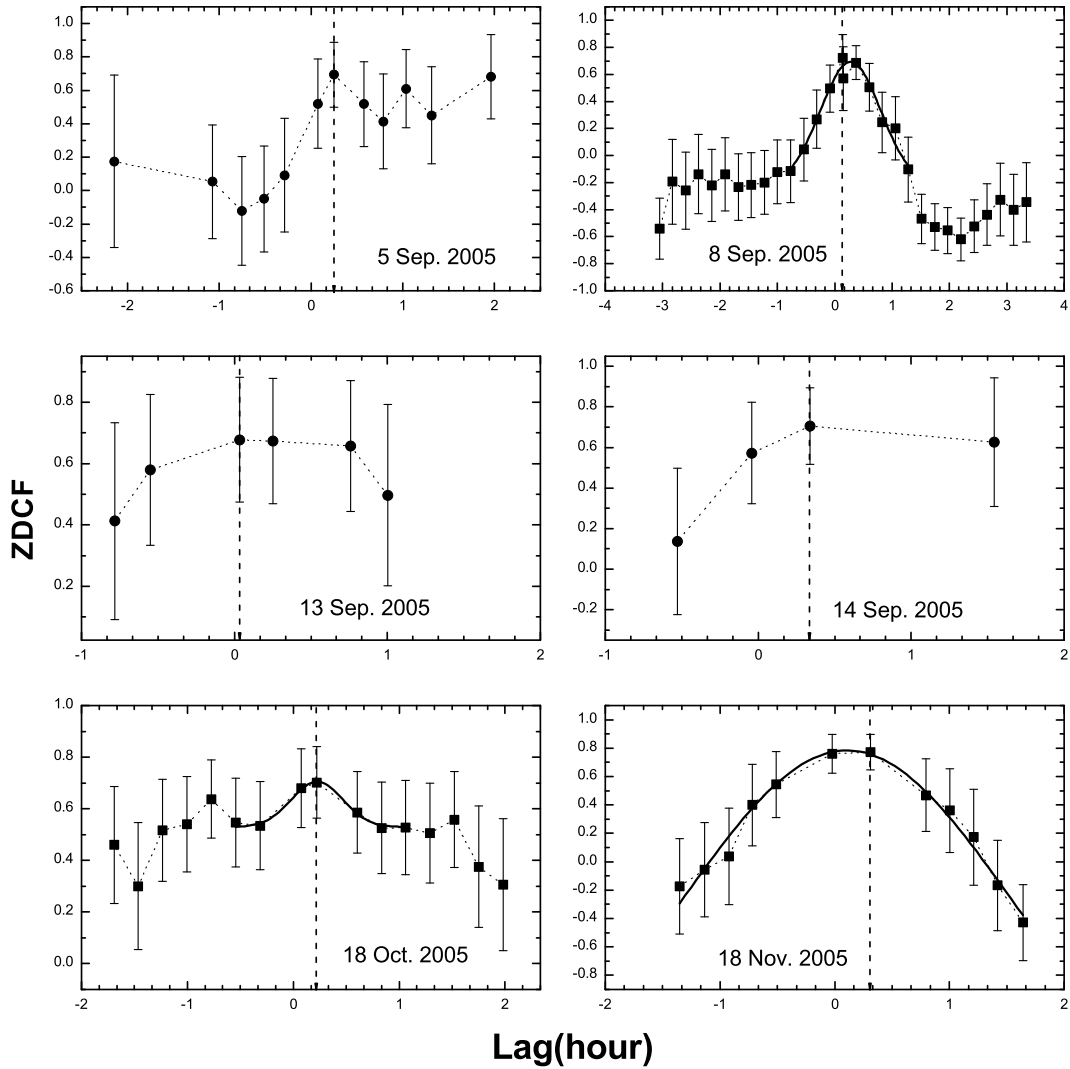}
      \caption{ZDCFs of the light curves in bands \emph{e} and \emph{m} on six variable
      nights. Dash lines show the maximum of the ZDCFs. Solid lines are the Gaussian fit lines.
              }
         \label{timelag}
   \end{figure}

It can be noticed that the shape of the light curves in three
bands are very similar (see Figs.~\ref{lcall} and \ref{lcintra}).
To look for correlations and possible time lags between them, the
z-transform discrete correlation function (ZDCF, Alexander
\citealp{alexander}) was applied to the light curves on ``variable
nights'' that were determined to have INOV. ZDCF is an improvement
of the discrete correlation function (DCF). ZDCF applies z
transformation to the correlation coefficients and uses equal
population bins rather than the equal time bins in DCF
\citep{edelson}. The calculation of ZDCF requires that the light
curves must have at least 12 points (Alexander
\citealp{alexander}). The ZDCFs between the light curves in band
\emph{e} and \emph{m} are illustrated in Fig.~\ref{timelag}. The
ZDCFs in three nights without enough observation points are not
good. So we only use the good ZDCFs in three nights to obtain the
result. The time lag was calculated by three methods. $\tau_{p}$
is the time lag corresponding to the maximum of ZDCF, $\tau_{c}$
and $\tau_{g}$ are the time lags calculated as centroid
\citep[see][]{raiteri,raiteri2005} and by Gaussian fit,
respectively. They are listed in Table~\ref{lagvalue}. The columns
of Tabel~\ref{lagvalue} are the date of observation, the maximum
of ZDCF and the time lags by three methods followed by their
errors. The time lags reveal that the variations in band \emph{e}
lead the variations in band \emph{m}. The average of the time lags
by three methods is 0.194 hours (11.6 minutes). It is in excellent
agreement with the lag obtained by Papadakis et al.
\cite{papadakis} on Jul. 5, 2001, 0.23 hours between band \emph{B}
and \emph{I}. These two results are all larger than 3 minutes
obtained by Stalin et al. \cite{stalin} between band \emph{V} and
\emph{R} on Oct. 22, 2001. But the different separation of band
frequency must be considered. An upper limit to the possible time
delay (10 minutes) between band \emph{B} and \emph{I} of 0716+714
obtained by Villata et al. \cite{villata2000} suggests that our
result is reasonable. It should be treated with caution, since it
is close to the temporal resolution of observation in each band.

\begin{table}
\caption{The time lags between band e and m.} \label{lagvalue}
\centering
\begin{tabular}{c c c c c}
\hline\hline
Date &   Max   &  $\tau_{p}$    & $\tau_{c}$ & $\tau_{g}$ \\
 mm/dd &         &    (hour) & (hour)        & (hour)     \\
\hline
09/08 & 0.724 & $0.142^{+0.001}_{-0.001}$  & 0.236$^{+0.136}_{-0.091}$ & 0.276$\pm$0.082\\
10/18 & 0.736 & $0.146^{+0.002}_{-0.001} $  & 0.156$^{+0.064}_{-0.081}$ & 0.225$\pm$0.337\\
11/18 & 0.770 & $0.312^{+0.287}_{-0.168} $  & 0.147$^{+0.164}_{-0.167}$ & 0.107$\pm$0.095\\
\hline
\end{tabular}
\end{table}

\subsection{Spectral variability}

The relationship between the optical spectral variability and the
brightness variation was investigated by many authors for a number
of BL Lacs (see e.g. Speziali \& Natali \citealp{speziali};
Papadakis et al. \citealp{papadakis}; \citealp{raiteri}; Villata
et al. \citealp{villata2004a}; \citealp{wu}; Stalin et al.
\citealp{stalin}). D'Amicis et al. \citep{damicis} reported that
the spectra of all their 8 BL Lac samples became bluer when they
got brighter. Similar results were obtained by Fiorucci et al.
\cite{fiorucci}. However Ramirez et al \citep{ramirez} showed that
the spectrum of a FSRQ PKS 0736+017 became red with increased
brightness. Gu et al. \citep{gu} found that all their 5 BL Lacs
tended to be bluer as they turned brighter, while two out of three
FSRQs tended to be redder when they were brighter. Villata et al.
\citep{villata2006} reported that FSRQs 3C 454.3 generally had the
redder-when-brighter behaviour during the 2004-2005 outburst. The
variation rate between the colour index (or spectral index) and
the brightness of each source studied above was different.

   \begin{figure}
   \centering
   \includegraphics[width=\columnwidth]{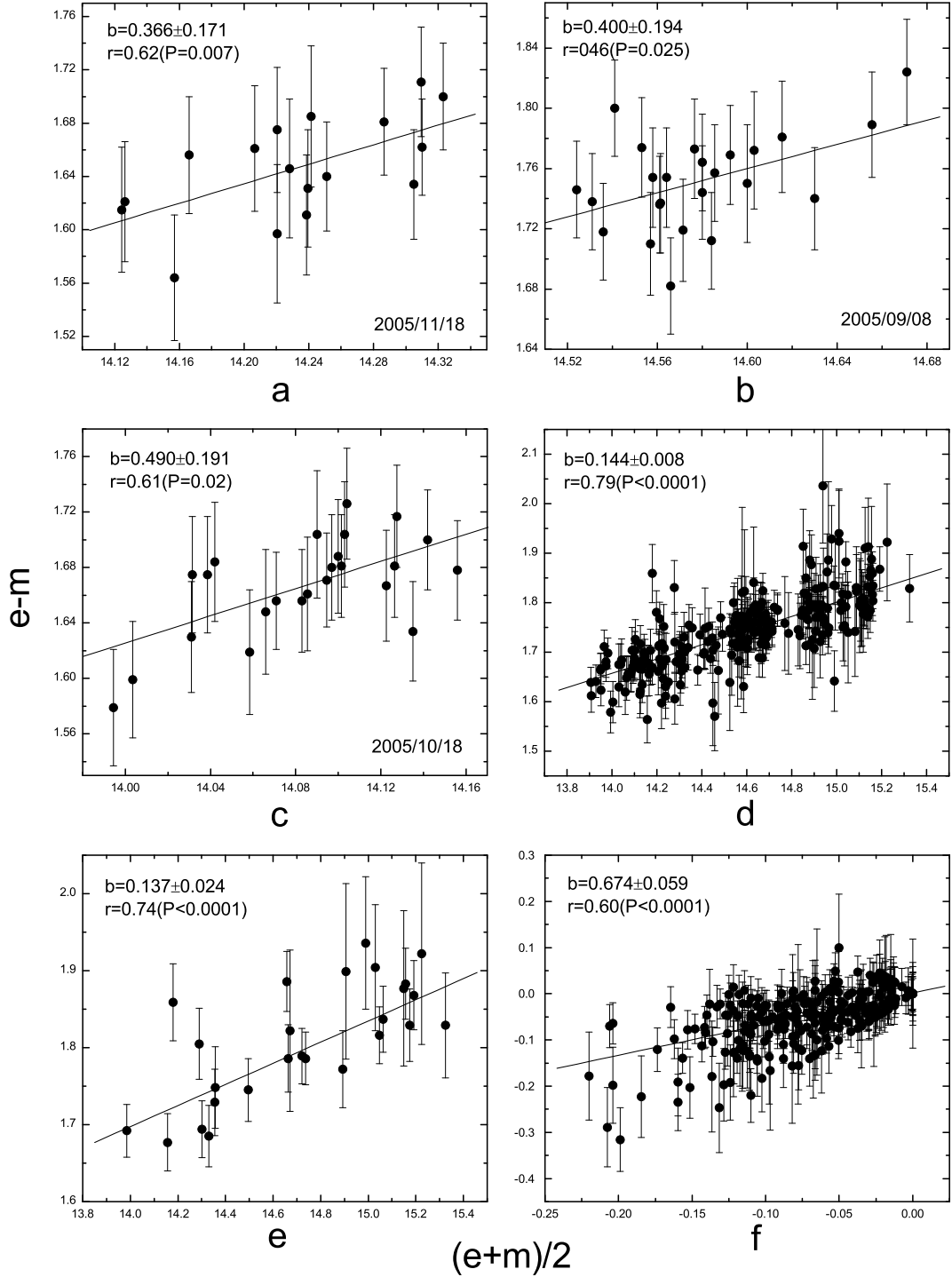}
      \caption{The relationship between colour index $(e-m)$ and
      source brightness $((e+m)/2)$. Solid lines are the best
      fittings to the data points. The fitted slope followed by its error is
indicated with \emph{b}, $r$ indicates the linear Pearson
correlation coefficient of the best fit, and $P$ is the
probability of $r=0$. Figs.~a, b and c --- for intranight data on
Nov. 18, 2005, Sep. 8, 2005 and Oct. 18, 2005, respectively;
Fig.~d --- for all data; Fig.~e --- for internight component;
Fig.~f --- for intranight component.
              }
         \label{color}
   \end{figure}

Based on our high temporal resolution observations, the spectral
variability with the brightness was investigated in this section.
The colour index $e-m$, which was calculated by coupling the BATC
\emph{e} and \emph{m} magnitudes taken within 20 minutes, was
adopted to denote the optical spectral slope. We took $(e+m)/2$ to
denote the brightness of the source. In our analysis, the
contribution of the host galaxy was not subtracted from the total
flux densities. But the results are not strongly affected by the
host galaxy, since it has colours similar to the AGN.
Figs.~\ref{color}.~a--c display the relationship between colour
index and brightness on three ``variable nights''. Only the most
densely observed three nights were plotted as examples, similar
results can be obtained from the data on other ``variable
nights''. Solid lines are the best fittings to the data points.
\emph{b} is the fitted slope followed by its error.
Fig.~\ref{color}.~d gave the relationship between the obtained 258
colour indices and their brightness during our monitoring. One can
obviously see that the values of $b$ for intranight points were
larger than that in Fig.~\ref{color}.~d. It suggests that the
spectral variability of the intranight fast flares and internight
variability may be different. To decompose the intranight
variation from the internight variation, we used the faintest
magnitudes in bands \emph{e} and \emph{m} on each observation
night to indicate the internight long term variation. Then the
intranight variations were obtained by subtracting the magnitudes
from the maximum for each night. The relationships between the
colour index and the brightness were drawn in Figs.~\ref{color}.~e
and f for the internight variations and the intranight fast
flares, respectively. The linear correlations are all significant
at a 0.95 confidence level, while the slope of fast flares
($0.674\pm0.059$) is larger than that of internight variations
($0.137\pm0.024$). So the spectral variability reveals that the
variability has ``strongly chromatic'' intranight fast flares and
``mildly chromatic'' internight variations. Our result is
consistent with that of Villata et al. \cite{villata2004a}. The
slopes in this work are larger than the slopes by Villata et al.
\citep{villata2004a}, since the considered frequency separation in
this work is larger than that in Villata et al.
\citep{villata2004a}.

\section{Summary}
We monitored BL Lac with a high temporal resolution with BATC
telescope during the period from July 5, 2005 to November 18,
2005. Many fast variations superposed on the long term trend were
recorded. The rising phase of the second flare lasted at least 44
days. During this rising process, the variability amplitudes are
1.374, 1.252 and 1.174 mag in band \emph{e}, \emph{i} and
\emph{m}, respectively. 799 optical multi-band observations and
258 colour indices were obtained. By analyzing the INOV, we
conclude that for the intranight variability, the amplitude of
variability is larger at the shorter wavelength, the variation
rate is also larger at shorter wavelength and it is comparable in
the rising and decaying phases within each band. An average time
lag between bands \emph{e} and \emph{m} was obtained by the method
of ZDCF. It must be treated with caution, because it is close to
the temporal resolution of our observations. Different variation
rates in different bands and the time lag will lead to spectrum
variations. The variability of BL Lac can be interpreted as having
two components by the analysis of spectral variability with the
brightness: a ``strongly-chromatic'' intranight component and a
``mildly-chromatic'' internight component. Our results confirmed
the conclusions about the components of the variability from
Villata et al. \citep{villata2004a}.

In the generally accepted scenario, the non-thermal emission from
BL Lacs includes low frequency synchrotron radiation and high
frequency inverse-Compton radiation, which are produced by
relativistic electrons in a jet oriented at a small angle to the
line of sight. The jet is powered and accelerated by a
supermassive black hole surrounded by an accretion disk
\citep[e.g.][]{ghisellini}. Although the details are still under
debate, this scenario can well explain most of the observation
properties of BL Lacs \citep[see e.g.][]{bottcher2000}. The
combination of ``strongly-chromatic'' intranight variability and
``mildly-chromatic'' internight variability is possible induced by
both intrinsic mechanisms and geometrical effects. The
``mildly-chromatic'' component can be interpreted by the variation
of Doppler factor on a spectrum slightly deviating from a power
law (Villata et al. \citealp{villata2004a}). The direction of the
jet varies as the knots move relativistically on helical
trajectories within an small angle with respect to the observer.
The internight variations can be introduced by the variation of
the direction of jet adding the spectrum slightly deviating from
the power law. The ``strongly-chromatic'' component can be
explained by particle acceleration and propagation by shock-in-jet
events \citep[e.g.][]{mastichiadis}. The particles are accelerated
by the shocks advancing down the jet, the optical synchrotron
emission is enhanced due to the accelerated particles and the
over-dense magnetic field. This kind of shock-in-jet model will
lead to a bluer-when-brighter phenomenon \citep{marscher}, which
is the same as the spectral variability in this paper. The
intranight component can also be explained by a simple model
representing the variability of a synchrotron component
\citep{vagnetti}. So the variability of BL Lac may be the results
of both intrinsic mechanisms and geometric effects.

\section*{Acknowledgments}

We owe great thanks to all the BATC staffs who make great efforts
to this campaign. We are grateful to the referee for valuable
comments and detailed suggestions that have been adopted to
improve this paper very much. This research is supported by the
National Natural Science Foundation of China under Grants 10433010
and 10521001. This work is partly supported by the Youth Growth
Foundation of Shandong University at Weihai.

\bsp \label{lastpage}

\end{document}